\documentclass[preprint,12pt]{elsarticle}

\usepackage{graphicx}
\usepackage{color,amsmath,amssymb,amsfonts}
\usepackage{bm}
\usepackage[caption=false]{subfig}

\def\be{\begin{equation}}
\def\ee{\end{equation}}
\def\ba{\begin{eqnarray}}
\def\ea{\end{eqnarray}}
\def\bd{\begin{displaymath}}
\def\ed{\end{displaymath}}
\def\bq{\begin{eqnarray}}
\def\eq{\end{eqnarray}}

\journal{Annals of Physics}

\begin{document}

\begin{frontmatter}

\title{Weak measurement with a coherent state pointer and its implementation
in optomechanical system}

\author[label1]{Gang Li\corref{corresp}} \ead{ligang0311@sina.cn}
\author[label2]{Ming-Yong Ye}
\author[label2]{Xiu-Min Lin}
\author[label1]{He-Shan Song\corref{corresp}} \ead{hssong@dlut.edu.cn}

\cortext[corresp]{Corresponding author}

\address[label1]{School of Physics and Optoelectronic Technology, \\
Dalian University of Technology, Dalian 116024 China}

\address[label2]{Fujian Provincial Key Laboratory of Quantum Manipulation and New
Energy Materials, \\
College of Physics and Energy, Fujian Normal University, Fuzhou 350007, China}

\begin{abstract}
Weak measurement with a coherent state pointer and in combination with an
orthogonal postselection can lead to a surprising amplification effect, and
we give a fire-new physical mechanism about the weak measurement in order to
understand this effect. Moreover, this physical mechanism is a general
result and based on it, we present a scheme of optomechanical system to
implement weak measurement amplification on an orthogonal postselection.
\end{abstract}



\end{frontmatter}



\section{Introduction}
\label{sec1}

Weak measurement was proposed by Aharonov, Albert and Vaidman in 1988 \cite%
{Aharonov88}. Initially the concept of weak measurement gives rise to some
controversy, but soon its physical meaning was clarified \cite{Duck89}.
Today weak measurement becomes one of the most prospective research tools
and it has been applied to various context including foundational questions
in quantum mechanics \cite{Lundden09,Yokota09} and some counter-intuitive
quantum paradoxes \cite{Aharonov05}. In weak measurement, a pointer is
weakly coupled to the system to be measured. Contrast to the project quantum
measurement, the output signal of the pointer in a weak measurement can be
far beyond the range of the eigenvalues of the system's observable if the
width of the initial wave function of the pointer is quite large. Such a
result is attributed to the interference of two or more slightly displaced
Gaussian states after a near-orthogonal postselection on the measured system
\cite{Aharonov88,Simon11}. Most recently it is revealed that weak
measurement can help to measure small physical quantities \cite{Bamber11} or
enable sensitive estimation of small physical parameters \cite%
{Hosten08,Howell09} that are difficult to be directly detected by
conventional techniques. Most of these weak measurement can be understood
using classical wave mechanics \cite{Ritchie91,Hosten08,Howell09} with one
exception in \cite{Pryde05}. Reviews of this filed can be found in \cite%
{Kofman12,Dressel14}.

Although weak measurement has many applications, its application in
optomechanics \cite{Girvin09,Marquardt13} is seldom investigated.
Optomechanical system consists of an optical cavity and a movable mirror.
The photons in the cavity can give rise to a radiation pressure on the
mirror and make it produce a displacement. In the weak coupling and when
there is only one photon in the cavity, the displacement of the mirror is
hard to be detected. The reason is that the displacement of the mirror
caused by one photon is much smaller than the width of the mirror wave
packet. Recently the standard scenario of weak measurement in \cite%
{Aharonov88} can be moved to Fock state space and understood by Fock-state
view \cite{Simon11}. And because of this, some of us showed that the
displacement of the mirror caused by one photon can be amplified if a weak
measurement is used and the initial state of the mirror is assumed to be in
the ground state \cite{Li14}. The result is obtained by retaining the Kerr
phase in the Ref. \cite{Bouwmeester12}.

In the most discussions about the standard weak measurements, the initial
state of the pointer is assumed to be the ground state and the interesting
features of weak measurements are usually due to the superposition of
several pointer states after a near-orthogonal postselection on the measured
system which can produce the amplification effect. The relative phases
between these pointer states play a key role and they can be adjusted
through the postselection (see Appendix A). As we know that coherent states
are regarded as classical-like states, naturally we want to know in weak
measurement whether there are some new features when the pointer is
initially prepared in coherent states. In the present paper, we will first
give a general discussion about weak measurements that use a coherent state
as the initial state of the pointer. It will be shown that there is a
fire-new physical mechanism for weak measurements that have an initially
coherent state pointer. It is regarded as a new mechanism because the
relative phases between the pointer states after the postselection, i.e.,
coherent quantum superposition \cite{French78}, can be due to the
noncommutativity of quantum mechanics induced by a coherent state pointer,
which is different from the standard weak measurement \cite%
{Aharonov88,Simon11} where the relative phases are prepared through the
postselection (see Appendix A), and this physical mechanism is a general
result for weak measurement with a coherent state pointer.

As an example we will consider a weak measurement with a coherent state
pointer in optomechanics. The weak measurement we will consider use the same
optomechanical model in Ref. \cite{Li14}
and the mirror amplification is discussed. We find there are following new
features about the obtained amplification with a coherent state pointer. (1)
The maximal amplification of the displacement of the mirror can reach the
level of the ground-state fluctuation and occur at time near zero, which is
very important for bad optomechanical cavities, i.e., sideband resolution is
small, and because of this, the implementation of our scheme is feasible in
experiment. (2) The relative phase between two mirror states after an orthogonal in this
paper is caused by the noncommutativity of quantum mechanics and the
relative phase in Ref. \cite{Li14} is caused by the Kerr phase generated by
the evolution of the Hamiltonian \cite{Mancini97,Bose97}. Therefore, the
generation mechanisms of the relative phases in this paper and the Ref. \cite%
{Li14} are different. Moreover, this conclusion in the Ref. \cite{Li14} is a
special case of the weak measurement amplification since it can only appear
in the optomechanical system.

The structure of our paper is as follows. In Sec. II, we give a general
discussion about weak measurement with a coherent state pointer. In Sec.
III, we present a scheme for implementation of weak measurement with a
coherent state pointer in optomechanics, and In Sec. IV, we
give the conclusion about the work.

\section{Weak measurement amplification with a coherent state pointer}
\label{sec2}

In the standard scenario of weak measurement, the system to be measured is
usually a two-level system and the pointer is a continuous system. The
Hamiltonian between the pointer and the system is given in general as
\begin{equation}
\hat{H}=\hbar \chi (t)\hat{\sigma}_{z}\otimes \hat{p},  \label{aa}
\end{equation}
where $\sigma_{z}$ is an observable of the system to be measured, $\hat{p}$
is the momentum operator of the pointer and $\chi (t)$ is a narrow pulse
function with integration $\chi $. Suppose $\hat{q}$ is position operator of
the pointer that is conjugates to $\hat{p}$, therefore there is $%
[q,p]=i\hbar $. As in Ref. \cite{Simon11}, if defining an annihilation
operator $\hat{c}=\frac{1}{2\sigma }\hat{q}+i\frac{\sigma }{ \hbar }\hat{p}$%
, where $\sigma $ is the zero-point fluctuation of the pointer ground state,
the Hamiltonian of Eq. (\ref{aa}) can be rewritten as
\begin{equation}
\hat{H}=-i\frac{\hbar ^{2}\chi (t)}{2\sigma }\hat{\sigma}_{z}(\hat{c}-\hat{c}
^{\dagger}).  \label{bb}
\end{equation}
Here we further considered the initial pointer state is a coherent state $%
|\alpha \rangle $ instead of the ground state $|0\rangle _{m}$ in \cite%
{Simon11}. Suppose the state $|+\rangle =\frac{1}{\sqrt{2}}(|0\rangle
_{s}+|1\rangle _{s})$ is the initial state of the system, where $|0\rangle
_{s}$ and $|1\rangle _{s}$ is eigenstates of $\hat{\sigma}_{z}$. Then the
time evolution of the total system is given by
\begin{eqnarray}
e^{-\frac{i}{\hbar }\int \hat{H}dt}|+\rangle |\alpha \rangle &=&\exp [-\eta
\hat{\sigma}_{z}(\hat{c}-\hat{c}^{\dagger })]|+\rangle |\alpha \rangle
\nonumber \\
&=&\frac{1}{\sqrt{2}}(|0\rangle _{s}D(\eta )D(\alpha )|0\rangle _{m}  \nonumber\\
&+&|1\rangle _{s}D(-\eta )D(\alpha )|0\rangle _{m}),  \label{cc}
\end{eqnarray}
where $\eta =\frac{\hbar \chi }{2\sigma }$ assumed to be very small, $D(\eta
)=\exp [\eta \hat{c}^{\dagger }-\eta ^{\ast }\hat{c}]$ is the displacement
operator and we have used $|\alpha \rangle =D(\alpha )|0\rangle _{m}$. When
the orthogonal postselection $|-\rangle $ is performed for the system, i.e.,
$\langle -|+\rangle =0$, then the final state of the pointer becomes
\begin{equation}
|\psi \rangle _{m}=\frac{1}{2}(D(\eta )D(\alpha )|0\rangle _{m}-D(-\eta
)D(\alpha )|0\rangle _{m})  \label{dd}
\end{equation}

For the sake of making the analysis simple, we can displace the state of Eq.
(\ref{dd}) to the origin point in phase space, defining $|\chi \rangle
_{m}=D^{\dag }(\alpha ))|\psi \rangle _{m}$ and there is
\begin{equation}
|\chi \rangle _{m}=\frac{1}{2}(e^{-i\varphi }D(\eta )|0\rangle
_{m}-e^{i\varphi }D(-\eta )|0\rangle _{m}),  \label{ee}
\end{equation}%
where the phases $e^{i\varphi }$ and $e^{-i\varphi }$ with $\varphi =-i\eta
(\alpha -\alpha ^{\ast })$ are obtained by using the property of the
displacement operators $D(\alpha )D(\beta )=\exp [\alpha \beta ^{\ast
}-\alpha ^{\ast }\beta ]D(\beta )D(\alpha )$, namely, the relative phase $%
e^{i2\varphi }$ is caused by the noncommutativity of quantum mechanics. In
other words, the relative phase $e^{i2\varphi }$ is induced by a coherent
state pointer since it is a quantum pointer. It is obvious that the
superposition state in Eq. (\ref{ee}) is known as "Coherent Quantum
Superposition" \cite{French78} since there is a relative phase.

For a weak measurement, the pointer states in Eq. (\ref{ee}) overlap
significantly. In order to detailedly observe overlap of the pointer states
in Eq. (\ref{ee}) and when $\varphi \ll 1$ and $\eta \ll 1$, we can then
perform a small quantity expansion about $\eta $ and $\varphi $ till the
first order, then
\begin{eqnarray}
|\chi \rangle _{m} &\approx &\frac{1}{2}((1-i\varphi )[1-\eta (\hat{c}-\hat{c%
}^{\dagger })]|0\rangle _{m}  \nonumber \\
&-&(1+i\varphi )[1+\eta (\hat{c}-\hat{c}^{\dagger })]|0\rangle _{m})  \nonumber\\
&\approx &-i\varphi |0\rangle _{m}+\eta |1\rangle _{m}.  \label{ff}
\end{eqnarray}%
This show that the unique advantage of the amplification using coherent
state pointer in weak coupling regime, where the relative phases caused by
the noncommutativity of quantum mechanics is achievable for the supposition
of $|0\rangle _{m}$ and $|1\rangle _{m}$ of the pointer on orthogonal
postselection, in sharp contrast to the fact that only $|1\rangle _{m}$ of
the pointer is generated using the ground state pointer on orthogonal
postselection \cite{Aharonov88,Simon11} (see Appendix A). Therefore, from
the origin of the relative phase and the supposition state of Eq. (\ref{ff})
induced by the relative phase, we can see clearly that the measuring device
(the pointer) with a coherent state have an effect on the postselection for
the system in weak measurement, which is not studied before.

The average displacement of the pointer position $\hat{q}$ is
\begin{equation}
\langle \hat{q}\rangle =\frac{\langle \chi |_{m}\hat{q}|\chi \rangle _{m}}{%
\langle \chi |\chi \rangle _{m}}-\langle 0|\hat{q}|0\rangle _{m}.  \label{hh}
\end{equation}%
In a similar way, the average displacement of the pointer momentum $\hat{p}$
is
\begin{equation}
\langle \hat{p}\rangle =\frac{\langle \chi |_{m}\hat{p}|\chi \rangle _{m}}{
\langle \chi |\chi \rangle _{m}}-\langle 0|\hat{p}|0\rangle _{m}.
\label{hhhpp}
\end{equation}%
So in this case of the orthogonal postselection, i.e., $\langle -|+\rangle
=0 $, we can find that
\begin{equation}
\langle \hat{q}\rangle =0.  \label{hh11}
\end{equation}%
and
\begin{equation}
\langle \hat{p}\rangle =\frac{\hbar }{2\sigma }\frac{2\varphi \eta }{\varphi
^{2}+\eta ^{2}}.  \label{hhhpp22}
\end{equation}

From Eq. (\ref{hh11}) and Eq. (\ref{hhhpp22}), it can be seen that $\langle
\hat{q}\rangle $ is zero in position space and $\langle \hat{p}\rangle $ is
non-zero in momentum space since $|\chi \rangle _{m}$ is a supposition
state. Such a result originates from non-zero relative phase $e^{i2\varphi }$
in Eq. (\ref{ee}) caused by the noncommutativity of quantum mechanics. We
note that when $\alpha $ is a real number there is $\varphi =0$ and the
displacement of the pointer momentum is zero, which is the same result as a
ground state pointer on non-orthogonal postselection \cite%
{Aharonov88,Simon11} (see Appendix A).

As is known to all, in standard weak measurement theory \cite%
{Aharonov88,Simon11}, the relative phase between two different pointer
states can be made through proper near-orthogonal postselection if the
pointer is initially assumed to be in the ground state (see Appendix A), but
this conclusion is not applicable to the condition of orthogonal
postselection and in this case the displecement of the pointer is zero.
However, our result shows that when the pointer is initially prepared in a
coherent state, it is after an orthogonal postselection that the relative
phase between two different pointer states, i.e., coherent quantum
superposition, can be caused by the noncommutativity of quantum mechanics.
It is because of the relative phase caused by the noncommutativity of
quantum mechanics that the interference of two different pointer states
after an orthogonal postselection can produce a large displacement of the
pointer. Therefore this perspective provided here show us a fire-new
physical mechanism about the weak measurement compared to the standard weak
measurement theory \cite{Aharonov88,Simon11} and this physical mechanism is
a general result for coherent state pointer.

\section{Implementation in optomechanics}

In the following we will consider an weak measurement model in
optomechanical system where the initial state of the pointer (a mirror) is
prepared in a coherent state.

\subsection{The optomechanical model}
\label{sec21}
We first consider the optomechanical system shown in Fig. 1(a), which
evolves under the following Hamiltonian \cite{Mancini97,Bose97}
\begin{equation}
H=\hbar \omega _{0}a^{\dagger }a+\hbar \omega _{m}c^{\dagger }c-\hbar
ga^{\dagger }a(c^{\dagger }+c),  \label{a}
\end{equation}%
where $\hbar $ is Plank's constant, $\omega _{0}$ and $a$ are angular
frequency and annihilation operator of the optical cavity mode,
respectively, $c$ is annihilation operator of mechanical system with angular
frequency $\omega _{m}$, and the optomechanical coupling strength $g=\frac{%
\omega _{0}}{L}\sigma $, where $L$ is the length of the cavity, $\sigma
=(\hbar /2m\omega _{m})^{1/2}$ is the zero-point fluctuation and $m$ is the
mass of mechanical system.

If the initial state of the mirror is prepared at the coherent state $%
|\alpha \rangle $ and one photon is input to the optomechahical cavity, then
the mirror state will evolve as follows \cite{Mancini97,Bose97}:
\begin{equation}
|\psi (\xi ,\varphi ,t)\rangle =e^{i\phi (t)}D(\xi (t))|\varphi (t)\rangle
_{m},  \label{aaa}
\end{equation}
where $e^{i\phi (t)}$ is the Kerr phase of one photon with $\phi
(t)=k^{2}(\omega _{m}t-\sin \omega _{m}t)$, $D(\xi (t))=\exp [\xi (t)c^{\dag
}-\xi ^{\ast }(t)c]$ is a displacement operator with $\xi
(t)=k(1-e^{-i\omega _{m}t})$, $\varphi (t)=\alpha e^{-i\omega _{m}t}$, and $%
k=g/\omega _{m}$ is the scaled coupling parameter. If no photon is input to
the optomechahical cavity, the mirror state will be the coherent state $%
|\varphi (t)\rangle _{m}$. The position displacement of the mirror caused by
one photon is $\langle \psi (\xi ,\varphi ,t)|\hat{q}|\psi (\xi ,\varphi
,t)\rangle -\langle \varphi (t)|\hat{q}|\varphi (t)\rangle $ with $\hat{q}$
being the position operator. It can be shown that the displacement can not
be bigger than $4k\sigma $ for any time $t$. Since $k=g/\omega _{m}$ can not
be bigger than $0.25$ in weak coupling condition \cite{Marshall03}, then the
position displacement of the mirror caused by one photon can not be bigger
than the zero-point fluctuation $\sigma $. From the literature \cite%
{Marshall03} we know that if the displacement of the mirror can be detected
experimentally it should be not smaller than $\sigma $. Therefore, the
displacement of the mirror caused by one photon can not be detected. In the
following we will show how the weak measurement with coherent state in
optomechanics can amplify the mirror's displacement.

\subsection{ The optomechanical model embedded in the interferometer}

Now consider a March-Zehnder interferometer which is shown in Fig. 1(b). The
optomechanical cavity A is embedded in one arm of the March-Zehnder
interferometer and a stationary Fabry-Pérot cavity B is placed in another
arm. The two beam splitters are both symmetric. The Hamiltonian of
optomechanical system is expressed as followed:
\begin{equation}
H=\hbar \omega _{0}(a^{\dagger }a+b^{\dagger }b)+\hbar \omega _{m}c^{\dagger
}c-\hbar ga^{\dagger }a(c^{\dagger }+c),  \label{bbb}
\end{equation}
where $b$ is annihilation operator of the optical cavity B. Other parameters
are the same as before. Here it is a weak measurement model where the mirror
is used as the pointer to measure the number of photon in cavity A, with $%
a^{\dag }a$ of the Eq. (\ref{bbb}) corresponding to $\hat{\sigma}_{z}$ in
Eq. (\ref{aa}) in the standard scenario of weak measurement and $c+c^{\dag}$
corresponds to $\hat{p}$.
\begin{figure}[tbp]
\centering
\includegraphics[scale=0.6]{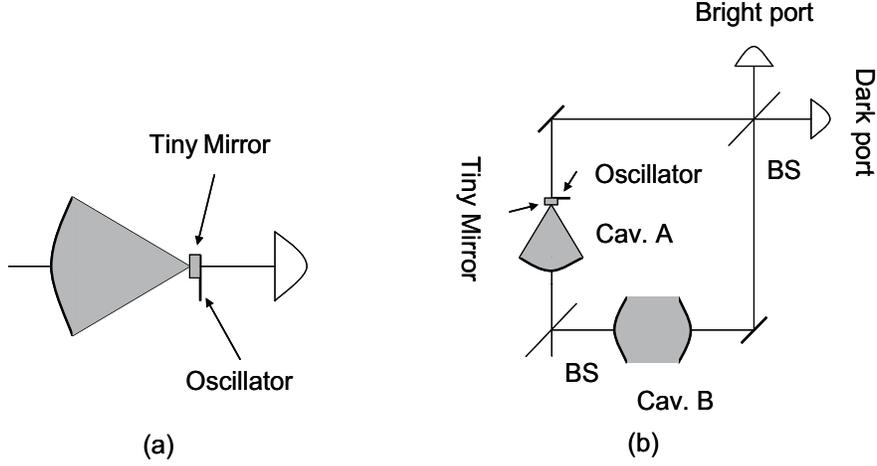}. \centering
\caption{(a) The optomechanical cavity. (b) The detection device with weak
measurement: the photon enters the first beam splitter, followed by an
optomechanical cavity A and a conventional cavity B. The photon weakly
excites the small mirror. After the second beam splitter, dark port is
detected, i.e., postselection is achieved in weak measurement where the
mirror has been excited by a photon, and fails otherwise.}
\end{figure}

Suppose that one photon is input into the interferometer, the state of the
photon after the first beam splitter becomes
\begin{equation}
|\psi_{i}\rangle =\frac{1}{\sqrt{2}}(|1\rangle _{A}|0\rangle _{B}+|0\rangle
_{A}|1\rangle _{B}),  \label{ccc}
\end{equation}
and after interacting weakly with the mirror prepared at coherent state $%
|\alpha \rangle $, $\alpha =|\alpha |e^{i\theta }$, where $|\alpha |$ and $%
\theta $ are real numbers called the amplitude and phase of the state,
respectively, the state of the total system is
\begin{eqnarray}
|\psi _{om}(t)\rangle &=&\frac{1}{\sqrt{2}}(|1\rangle _{A}|0\rangle
_{B}|\psi (\xi ,\varphi ,t)\rangle _{m}+|0\rangle _{A}  \nonumber \\
&&|1\rangle _{B}|\varphi (t)\rangle _{m}),  \label{ddd}
\end{eqnarray}
When the photon is detected at dark port, i.e., in the language of weak
measurement \cite{Li14} the postselected state of the single-photon is
\begin{equation}
|\psi _{f}\rangle =\frac{1}{\sqrt{2}}(|1\rangle _{A}|0\rangle _{B}-|0\rangle
_{A}|1\rangle _{B}),  \label{eee}
\end{equation}
which is orthogonal to $|\psi _{i}\rangle$. Then the final state of the
mirror becomes
\begin{equation}
|\Psi _{os}(t)\rangle =\frac{1}{2}(|\psi (\xi ,\varphi ,t)\rangle
_{m}-|\varphi (t)\rangle _{m}).  \label{fff}
\end{equation}

For the sake of making the analysis simple, we can displace the state of Eq.
(\ref{fff}) to the origin point in phase space, defining $|\chi
_{os}(t)\rangle =D^{\dag }(\varphi (t))|\Psi _{os}(t)\rangle $ and there is
\begin{equation}
|\chi _{os}(t)\rangle =\frac{1}{2}(\exp [i\phi (t)+i\phi (\alpha ,t)]|\xi
(t)\rangle _{m}-|0\rangle _{m}),  \label{ggg}
\end{equation}%
where the phase $e^{i\phi (\alpha ,t)}$ with $\phi (\alpha ,t)=-i[(\xi
(t)\varphi ^{\ast }(t)-\xi ^{\ast }(t)\varphi (t))]$ is a relative phase
between the coherent state $|\xi (t)\rangle $ and the ground state $%
|0\rangle $, and it is obtained through the property of the displacement
operators $D(\alpha )D(\beta )=\exp [\alpha \beta ^{\ast }-\alpha ^{\ast
}\beta ]D(\beta )D(\alpha )$, i.e., the noncommutativity of quantum
mechanics. And by the way, in the Ref. \cite{Li14} the relative phase
between two mirror states after an orthogonal postselection is caused by the
Kerr phase due to the evolution of the Hamiltonian \cite{Mancini97,Bose97}
and leads to the amplification. This can only appear in the optomechanical
system and it can be said to be a special case of the amplification. But the
amplification caused by the phase due to the noncommutativity of quantum
mechanics is a general result for coherent state pointer.

Next, we will show how the amplification of the displacement of the mirror
is generated via this relative phase $e^{i\phi (\alpha ,t)}$.

\begin{figure}[tbp]
\centering
\begin{tabular}{c c}
\captionsetup[subfigure]{margin={0.7cm,0cm}} \subfloat[]{%
\includegraphics[scale=0.42]{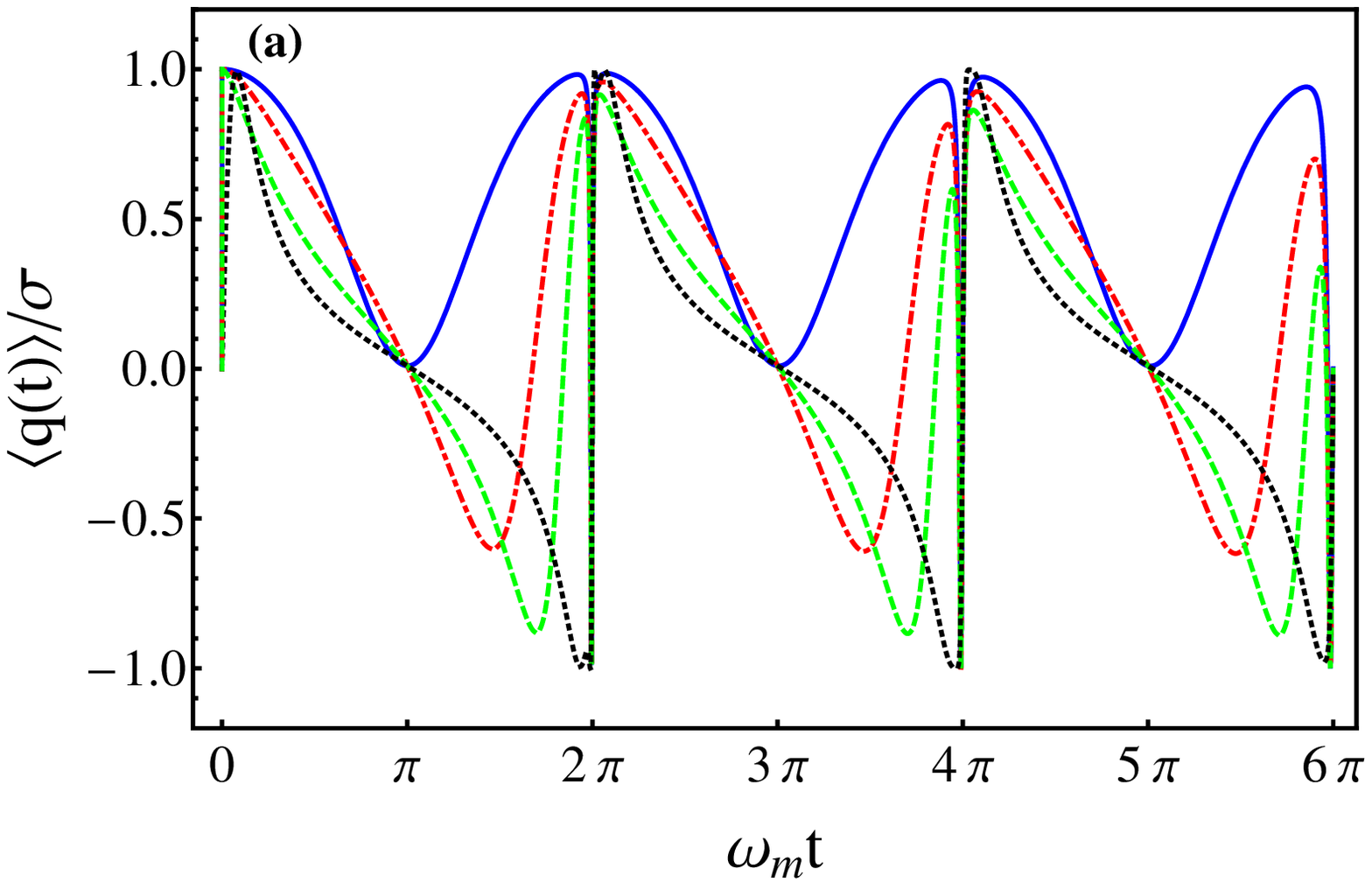}} & %
\captionsetup[subfigure]{margin={0.7cm,0cm}} \subfloat[]{%
\includegraphics[scale=0.42]{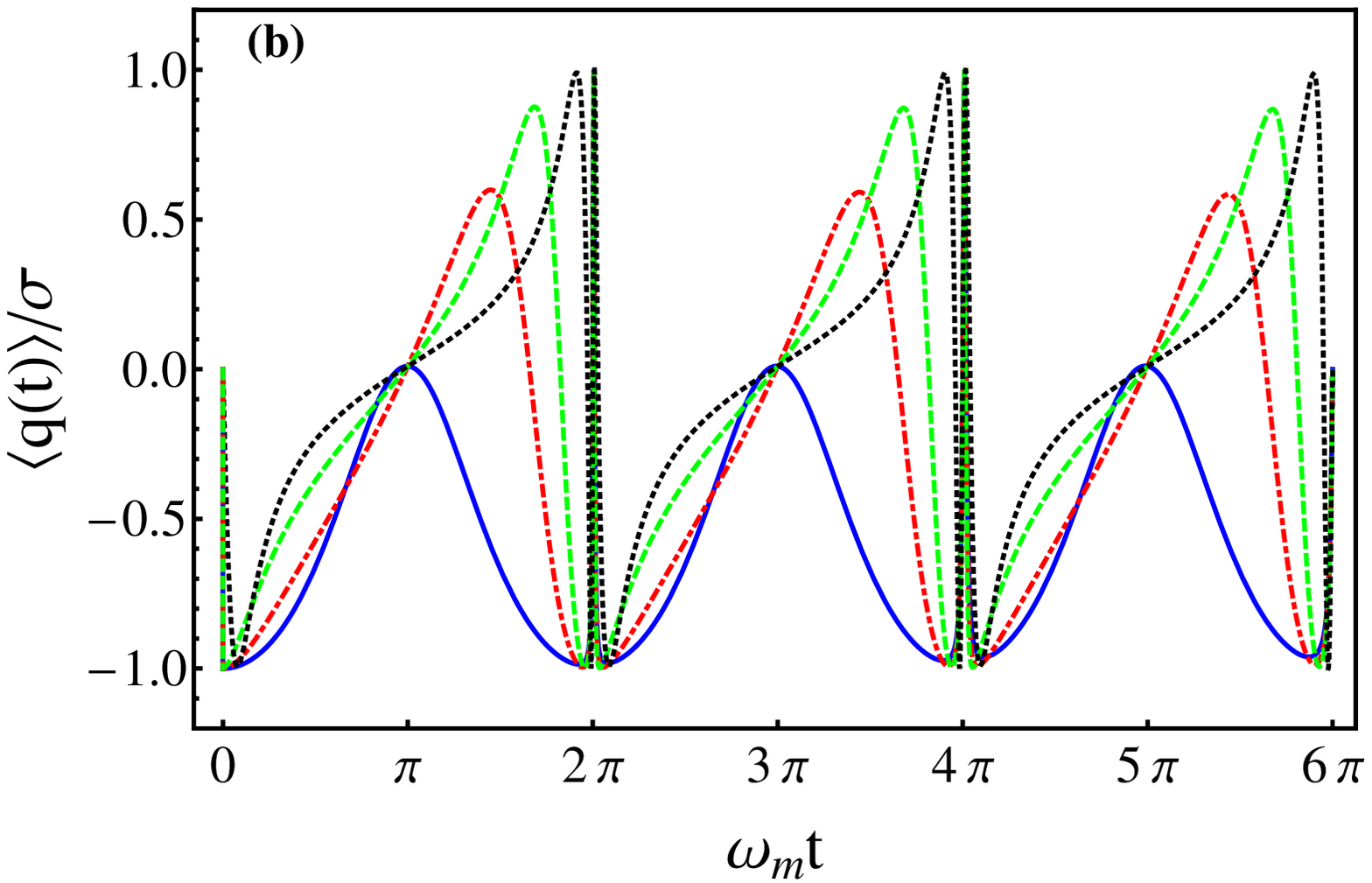}}%
\end{tabular}
\centering
\caption{(a) The average displacement $\langle q(t)\rangle/\protect\sigma$
with $k=0.005$ for different amplitudes $|\protect\alpha|$ and different
phases $\protect\theta$ [$|\protect\alpha|=1/2$ and $\protect\theta=0$
(blue-solid line), $|\protect\alpha|=1$ and $\protect\theta=\protect\pi/3$
(red-dotted-dashed line), $|\protect\alpha|=2$ and $\protect\theta=5\protect%
\pi/12$ (green-dashed line), and $|\protect\alpha|=4$ and $\protect\theta=%
\protect\pi/2$ (black-dotted line)]. (b) The average displacement $\langle
q(t)\rangle/\protect\sigma$ with $k=0.005$ for different amplitudes $|%
\protect\alpha|$ and different phases $\protect\theta$ [$|\protect\alpha%
|=1/2 $ and $\protect\theta=\protect\pi$ (blue-solid line), $|\protect\alpha%
|=1$ and $\protect\theta=4\protect\pi/3$ (red-dotted-dashed line), $|\protect%
\alpha|=2$ and $\protect\theta=17 \protect\pi/12$ (green-dashed line), and $|%
\protect\alpha|=4$ and $\protect\theta=3\protect\pi/2$ (black-dotted line)].}
\end{figure}

\subsection{ Amplification about position variable $q$ via coherent state
pointer}

The average displacement of pointer variable $\hat{q}$ of the mirror is
\begin{equation}
\langle \hat{q}\rangle =\frac{\langle \Psi _{os}(t)|\hat{q}|\Psi
_{os}(t)\rangle }{\langle \Psi _{os}(t)|\Psi _{os}(t)\rangle }-\langle
\varphi (t)|\hat{q}|\varphi (t)\rangle ,  \label{aaaa}
\end{equation}%
which will be
\begin{equation}
\langle \hat{q}\rangle =\frac{Tr(|\chi _{os}(t)\rangle \langle \chi _{os}(t)|%
\hat{q})}{Tr(|\chi _{os}(t)\rangle \langle \chi _{os}(t)|)}-Tr(0\rangle
_{m}\langle 0|_{m}\hat{q}).  \label{bbbb}
\end{equation}%
It is clearly that through Eq. (\ref{ggg}) we can obtain the average
displacement of the mirror's position $q$ and as a result we have
\begin{eqnarray}
\langle q(t)\rangle &=&\sigma \lbrack \xi (t)+\xi ^{\ast }(t)-e^{-\frac{|\xi
(t))|^{2}}{2}}(e^{i\phi (t)+i\phi (\alpha ,t)}\xi (t)  \nonumber \\
&+&e^{-(i\phi (t)+i\phi (\alpha ,t))}\xi ^{\ast }(t))]/[2-e^{-\frac{|\xi
(t)|^{2}}{2}}(e^{i\phi (t)}  \nonumber \\
&\times &e^{i\phi (\alpha ,t)}+e^{-(i\phi (t)+i\phi (\alpha ,t))})],
\label{cccc}
\end{eqnarray}%
where the phase $e^{i\phi (\alpha ,t)}$ is the relative phase from the
noncommutativity of quantum mechanics we mentioned in the previous
subsection.

The average displacement $\langle q(t)\rangle /\sigma $ of the mirror is
shown in Fig. 2 as a function of time $\omega _{m}t$ with $k=0.005$ for
different coherent states $|\alpha =|\alpha |e^{i\theta }\rangle $. It is
clear from Fig. 2 that the amplification occurring near $\omega _{m}t=0$ can
reach the strong-coupling limiting (the level of the ground-state
fluctuation) \cite{Marshall03} $\langle q\rangle =\sigma $ or $-\sigma $.
This result is very important for bad optomechanical cavities where its
decay rate is very large so that the photon will have a large probability to
leak out from the cavity within a very short time. And the time interval
where there are large amplification around $\omega _{m}t=0$ is broad for
some coherent states and therefore can be easier to be detected, which is
contrast to the time interval of amplification around $\omega _{m}t=0$ in
Ref. \cite{Li14} where it is very narrow. Note that the maximal displacement
of the mirror caused by one photon in the cavity (see Fig. 1(a)) is $%
4k\sigma $ and the displacement achieved here can be $\sigma $ or $-\sigma $
using weak measurement, therefore the amplification factor can be $Q=\pm
1/4k $ which is $\pm 50$ when $k=0.005$.

\subsection{ Small quantity expansion about time for amplification}

To study the amplification effects occurring near $\omega _{m}t=0$, for Eq. (%
\ref{ggg}) we can then make a small quantity expansion about time $0$ till
the second order. Suppose that $\omega _{m}t\ll 1$, and $k\ll 1$, then
\begin{eqnarray}
|\chi _{os}(t)\rangle &\approx &\frac{1}{2}[ik\omega _{m}t|1\rangle
_{m}+i2k|\alpha |(\frac{(\omega _{m}t)^{2}}{2}\sin \theta  \nonumber \\
&+&\omega _{m}t\cos \theta )|0\rangle _{m}],  \label{dddd}
\end{eqnarray}%
which is a superposition of $|0\rangle _{m}$ and $|1\rangle _{m}$ and the
amplitude of $|0\rangle _{m}$ is due to the relative phase $e^{i\phi (\alpha
,t)}$ caused by noncommutativity of quantum mechanics when we use a non-zero
coherent state ($|\alpha |\neq 0$). The superposition of $|0\rangle _{m}$
and $|1\rangle _{m}$ is the key to obtain amplification. From Appendix A, we
know that when
\begin{equation}
k\omega _{m}t=2k|\alpha |(\frac{(\omega _{m}t)^{2}}{2}\sin \theta +\omega
_{m}t\cos \theta )  \label{ffff}
\end{equation}%
the displacement of the mirror can reach the maximal value $\sigma $, and
when
\begin{equation}
k\omega _{m}t=-2k|\alpha |(\frac{(\omega _{m}t)^{2}}{2}\sin \theta +\omega
_{m}t\cos \theta )  \label{gggg}
\end{equation}%
the displacement of the mirror can reach the minimal value $-\sigma $.
Hence, the mirror state achieving the maximal positive amplification is $%
\frac{1}{\sqrt{2}}(|0\rangle _{m}+|1\rangle _{m})$ and the one achieving the
maximal negative amplification is $\frac{1}{\sqrt{2}}(|0\rangle
_{m}-|1\rangle _{m})$. We emphasis that the superposition of $|0\rangle _{m}$
and $|1\rangle _{m}$ achieved here is through an orthogonal postselection
that is impossible in the standard weak measurement with an orthogonal
postselection (see Appendix A). The superpostion achieved here is due to the
noncommutativity of quantum mechanics and the superpostion in the standard
weak measurement is due to the non-orthogonal postselection (see Appendix
A), therefore it is a new mechanism for weak measurement. Substituting Eq. (%
\ref{dddd}) into Eq. (\ref{bbbb}), the average value of displacement
operator $q$ is given by
\begin{eqnarray}
\langle q(t)\rangle _{\omega _{m}t\ll 1} &=&\sigma \lbrack 4k^{2}|\alpha
|((\omega _{m}t)^{2}\cos \theta +\frac{(\omega _{m}t)^{3}}{2}  \nonumber \\
&&\sin \theta ))]/[k^{2}(\omega _{m}t)^{2}+4k^{2}|\alpha |^{2}(\omega _{m}t
\nonumber \\
&&\cos \theta +\frac{(\omega _{m}t)^{2}}{2}\sin \theta )^{2}],  \label{eeee}
\end{eqnarray}%
which is plotted in Fig. 3 for different coherent state $|\alpha =|\alpha
|e^{i\theta }\rangle $ with $k=0.005$ and $\omega _{m}t=0.001$.
\begin{figure}[tbp]
\centering
\includegraphics[scale=0.5]{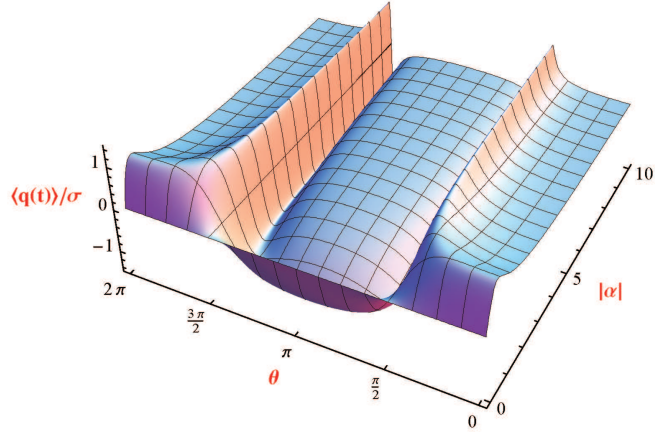}
\caption{The average displacement $\langle q(t)\rangle /\protect\sigma $ as
function of amplitude $|\protect\alpha |$ and the phase $\protect\theta $
with $k=0.005$ and time $\protect\omega _{m}t=0.001$.}
\end{figure}

\subsection{Amplification about position variable $q$ based on the
dissipation.}

Taking into account of dissipation, the master equation of the mechanical
system \cite{Bose97} is given by
\begin{eqnarray}
\frac{d\rho (t)}{dt} &=&-\frac{i}{\hbar }[H,\rho (t)]  \notag \\
&+&\frac{\gamma _{m}}{2}[{2c\rho (t)c^{\dag }-c^{\dag }c\rho (t)-\rho
(t)c^{\dag }c]},  \label{ii}
\end{eqnarray}%
where $\gamma _{m}$ is the damping constant. And the above master equation
have an analytical solution for the evolution of the mechanical system,
which is obtained by the method in Ref. \cite{Bose97}.

If steps corresponding to those of Eqs. (\ref{ccc})--(\ref{eee}) are
carefully carried out in this case, then the final state of mirror becomes
\begin{eqnarray}
\rho _{m}(t) &=&\frac{1}{4}[|\varphi _{1}(\gamma ,t)\rangle _{m}\langle
\varphi _{1}(\gamma ,t)|_{m}  \nonumber \\
&-&e^{i\phi (t)+i\phi (\alpha ,t)/2-D(\gamma ,t)}|\varphi _{1}(\gamma
,t)\rangle _{m}\langle \varphi _{0}(\gamma ,t)|_{m}  \nonumber \\
&-&e^{-i\phi (t)-i\phi (\alpha ,t)/2-D(\gamma ,t)}|\varphi _{0}(\gamma
,t)\rangle _{m}\langle \varphi _{1}(\gamma ,t)|_{m}  \nonumber \\
&+&|\varphi _{0}(\gamma ,t)\rangle _{m}\langle \varphi _{0}(\gamma ,t)|_{m}],
\label{jj}
\end{eqnarray}%
where $\gamma =\gamma _{m}/\omega _{m}$ and $\varphi _{n}(\gamma ,t)=\alpha
e^{-(i+\gamma /2)\omega _{m}t}+\frac{ikn}{i+\gamma /2}(1-e^{-(i+\gamma
/2)\omega _{m}t})$ $(n=0,1)$ are the amplitude of the coherent states of the
mirror and
\begin{eqnarray}
D(\gamma ,t) &=&\frac{k^{2}\gamma }{2(1+\gamma ^{2}/4)}[\omega _{m}t+\frac{%
1-e^{-\gamma \omega _{m}t}}{\gamma }  \nonumber \\
&-&\frac{e^{(i-\gamma /2)\omega _{m}t}-1}{i-\gamma /2}+\frac{e^{-(i-\gamma
/2)\omega _{m}t}-1}{i+\gamma /2}].  \label{kk}
\end{eqnarray}

The relative phase $e^{i\phi (t)+i\phi (\alpha ,t)/2}$ between the coherent
states $|\varphi _{1}(\gamma ,t)\rangle $ and $|\varphi _{0}(\gamma
,t)\rangle $ is composed of the Kerr phase $e^{i\phi (t)}$ and the phase $%
e^{i\phi (\alpha ,t)/2}$ obtained by interchanging two displacement
operators.

Substituting Eq. (\ref{jj}) into Eq. (\ref{bbbb}) using $\rho _{m}(t)$ and $%
|\varphi _{0}(\gamma ,t)\rangle $ instead of $|\chi _{os}(t)\rangle $ and $%
|0\rangle $, respectively. As a result, we have
\begin{eqnarray}
\langle q(t)\rangle &=&\sigma \lbrack \varphi _{1}(\gamma ,t)+\varphi
_{1}^{\ast }(\gamma ,t)-e^{-\frac{|\varphi _{1}(\gamma ,t)-\varphi
_{0}(\gamma ,t)|^{2}}{2}}  \nonumber \\
&\times &(e^{i\phi (t)+i\phi (\alpha ,t)/2+i\tau -D(\gamma ,t)}(\varphi
_{1}(\gamma ,t)+\varphi _{0}^{\ast }(\gamma ,t))  \nonumber \\
&+&e^{-(i\phi (t)+i\phi (\alpha ,t)/2+i\tau )-D(\gamma ,t)}(\varphi
_{1}^{\ast }(\gamma ,t)  \nonumber \\
&+&\varphi _{0}(\gamma ,t)))+\varphi _{0}(\gamma ,t)+\varphi _{0}^{\ast
}(\gamma ,t)]/[2  \nonumber \\
&-&e^{-\frac{|\varphi _{1}(\gamma ,t)-\varphi _{0}(\gamma ,t)|^{2}}{2}%
}(e^{i\phi (t)+i\phi (\alpha ,t)/2+i\tau -D(\gamma ,t)}  \nonumber \\
&+&e^{-(i\phi (t)+i\phi (\alpha ,t)/2)}e^{-i\tau -D(\gamma ,t)})],
\end{eqnarray}%
where $\tau ={{\mathop{\rm Re}\nolimits}{\varphi _{0}}(t){\mathop{\rm Im}%
\nolimits}{\varphi _{1}}(t)-{\mathop{\rm Im}\nolimits}{\varphi _{0}}(t){\ %
\mathop{\rm Re}\nolimits}{\varphi _{1}}(t)}$.
\begin{figure}[tbp]
\centering
\begin{tabular}{cc}
\captionsetup[subfigure]{margin={0.7cm,0cm}} \subfloat[]{
\includegraphics[scale=0.42]{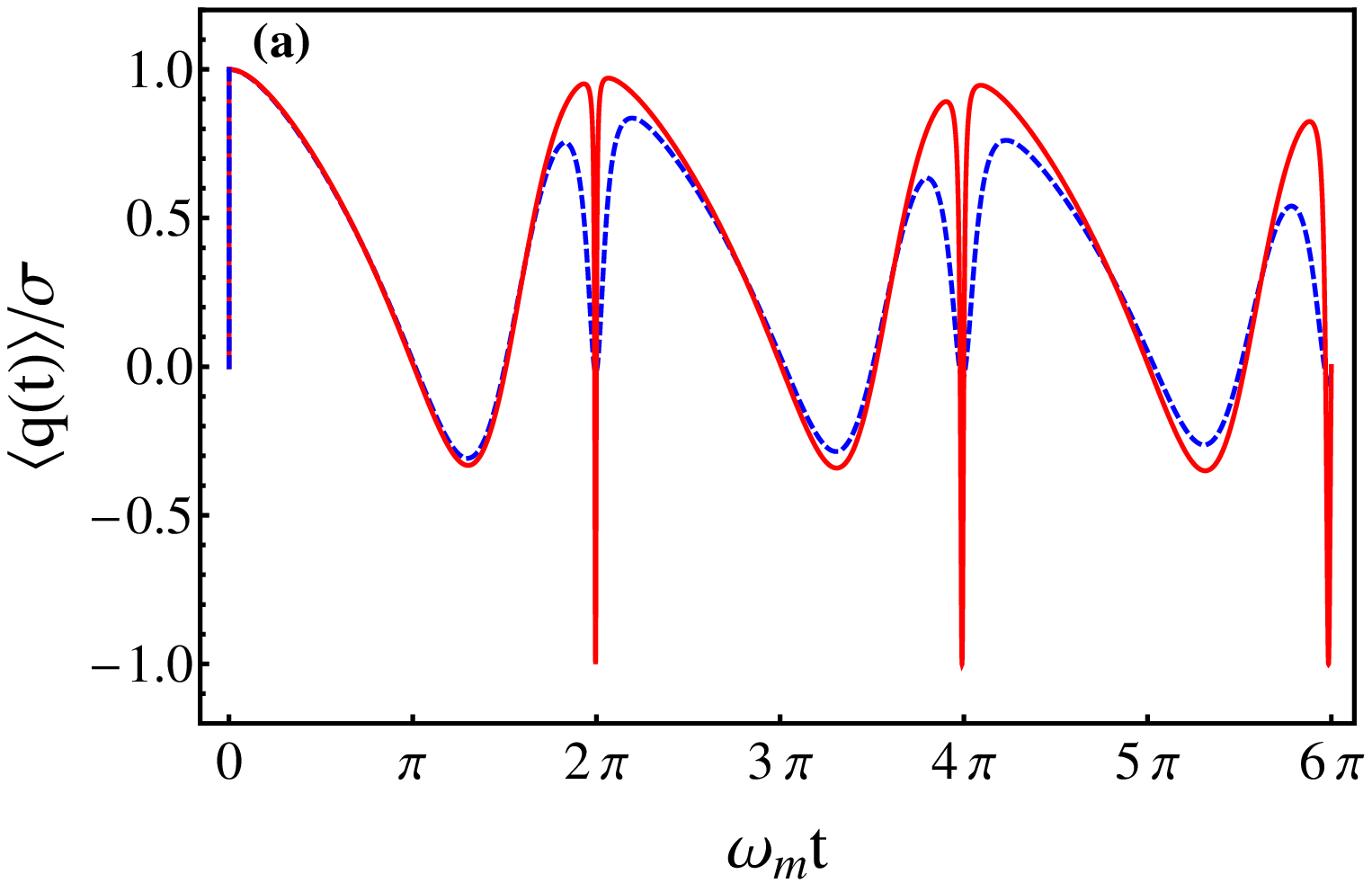}} & %
\captionsetup[subfigure]{margin={0.7cm,0cm}} \subfloat[]{%
\includegraphics[scale=0.42]{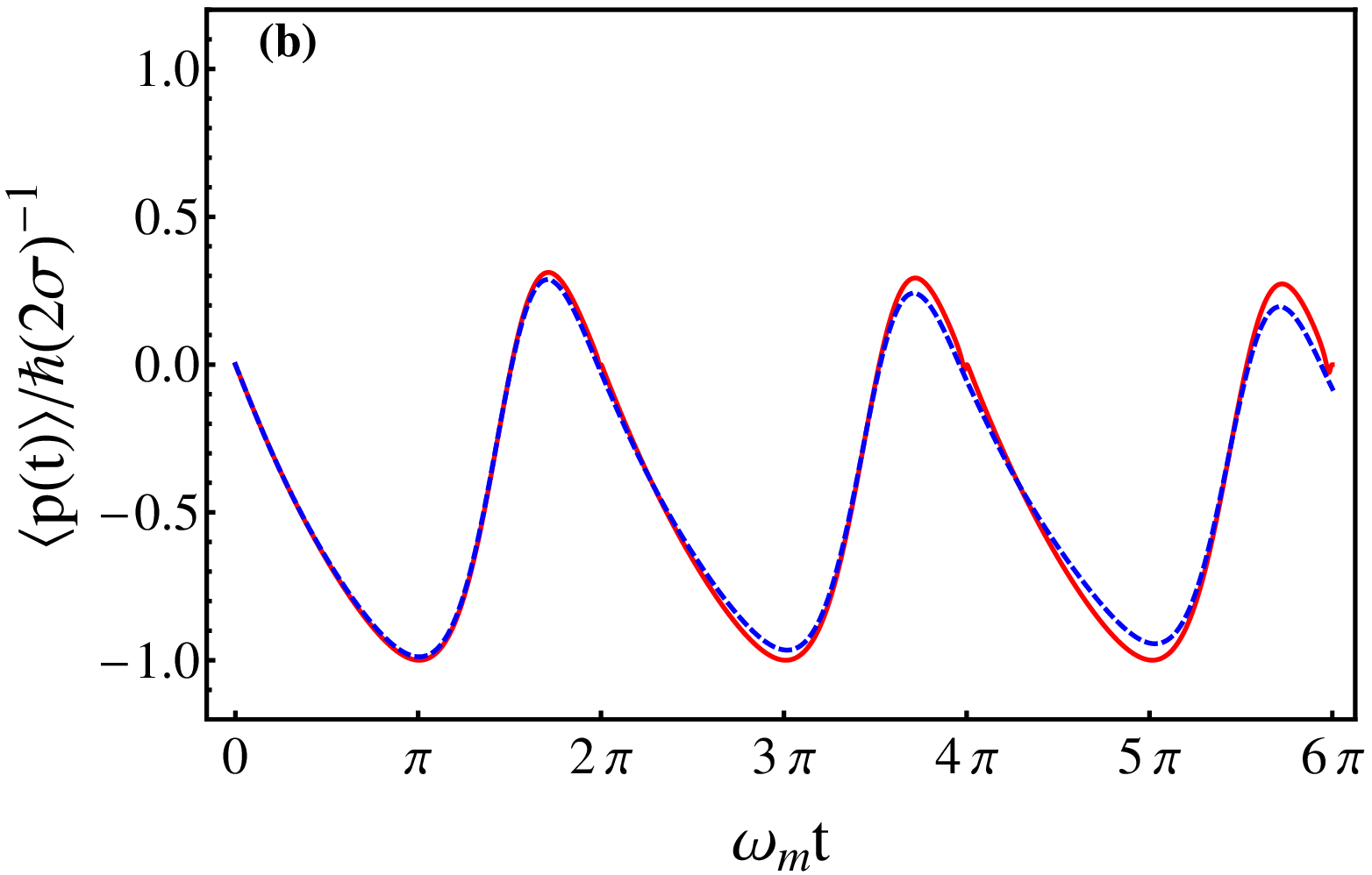}}%
\end{tabular}
\centering
\caption{For $|\protect\alpha |=\frac{1}{\protect\sqrt{2}}$ and $\protect%
\theta =\protect\pi /4$, (a) the average displacement $\langle q(t)\rangle /%
\protect\sigma $ is as a function of $\protect\omega _{m}t$ with $k=0.005$, $%
\protect\gamma =0$ (solid line) and $\protect\gamma =0.005$ (dashed line).
(b) the average displacement $\langle p(t)\rangle /\hbar (2\protect\sigma %
)^{-1}$ is as a function of $\protect\omega _{m}t$ with $k=0.005$, $\protect%
\gamma =0$ (solid line) and $\protect\gamma =0.005$ (dashed line)}
\end{figure}

The average displacement $\langle q(t)\rangle /\sigma $ of the mirror is
shown in Fig. 4(a) as a function of $\omega _{m}t$ with $k=0.005$, $\gamma
=0 $ (solid line) and $\gamma =0.005$ (dashed line). It can be seen from
Fig. 4(a) that all the amplification values in the presence of the damping
are reduced (dashed line), but the actual $\gamma $ can be very small ($%
\gamma =5\times 10^{-7}$ in \cite{Bouwmeester12}). The result for $\gamma
=5\times 10^{-7}$ is almost the same as the one for $\gamma =0$.

\subsection{ Amplification about momentum variable $p$ via coherent state
pointer}

Next, without loss of generality, we would like to discuss the amplification
of the momentum variable $p$ of the mirror in presence of damping.
\begin{figure}[tbp]
\centering
\begin{tabular}{cc}
\captionsetup[subfigure]{margin={0.7cm,0cm}} \subfloat[]{
\includegraphics[scale=0.42]{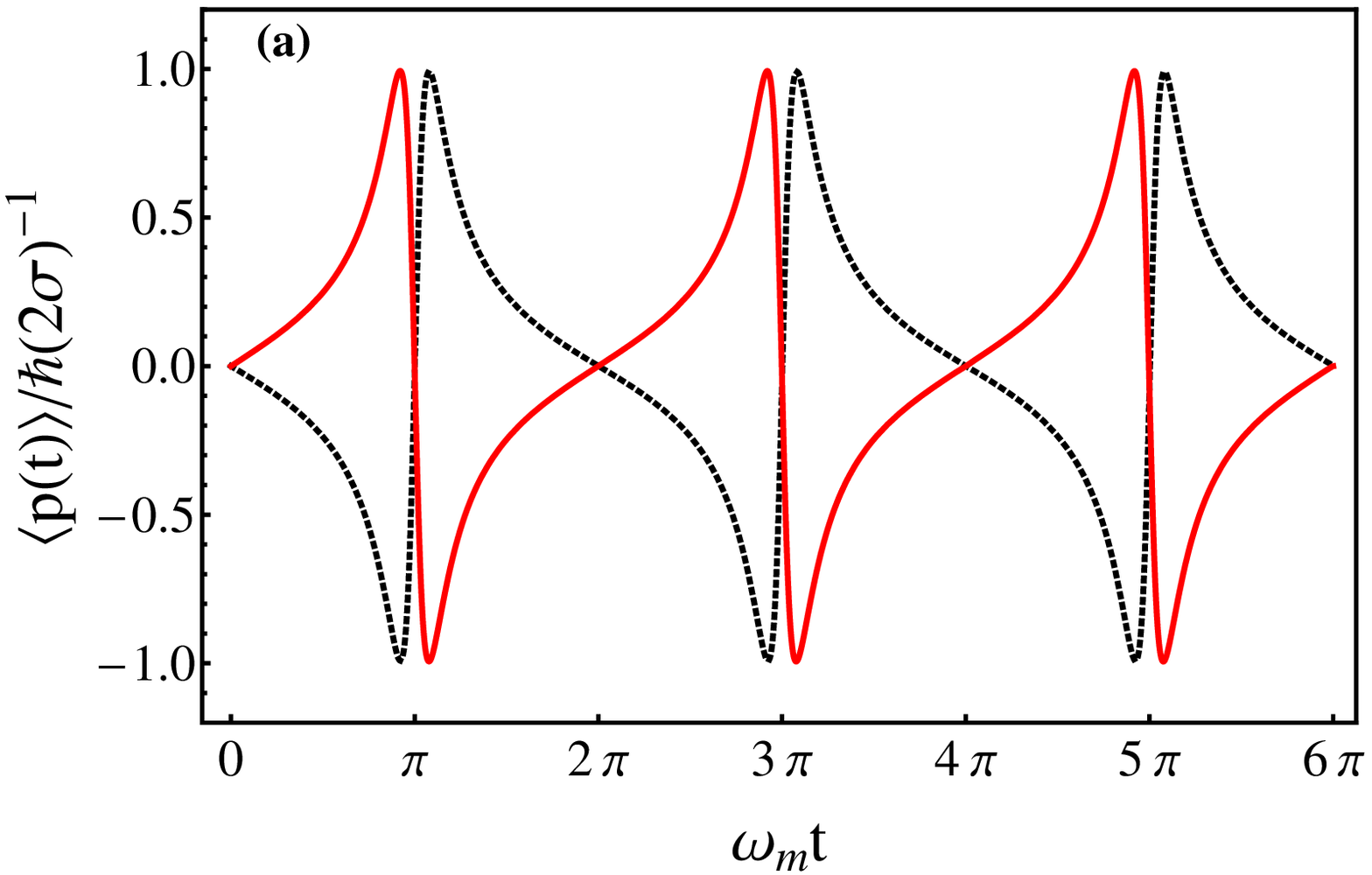}} &
\captionsetup[subfigure]{margin={0.7cm,0cm}} \subfloat[]{
\includegraphics[scale=0.42]{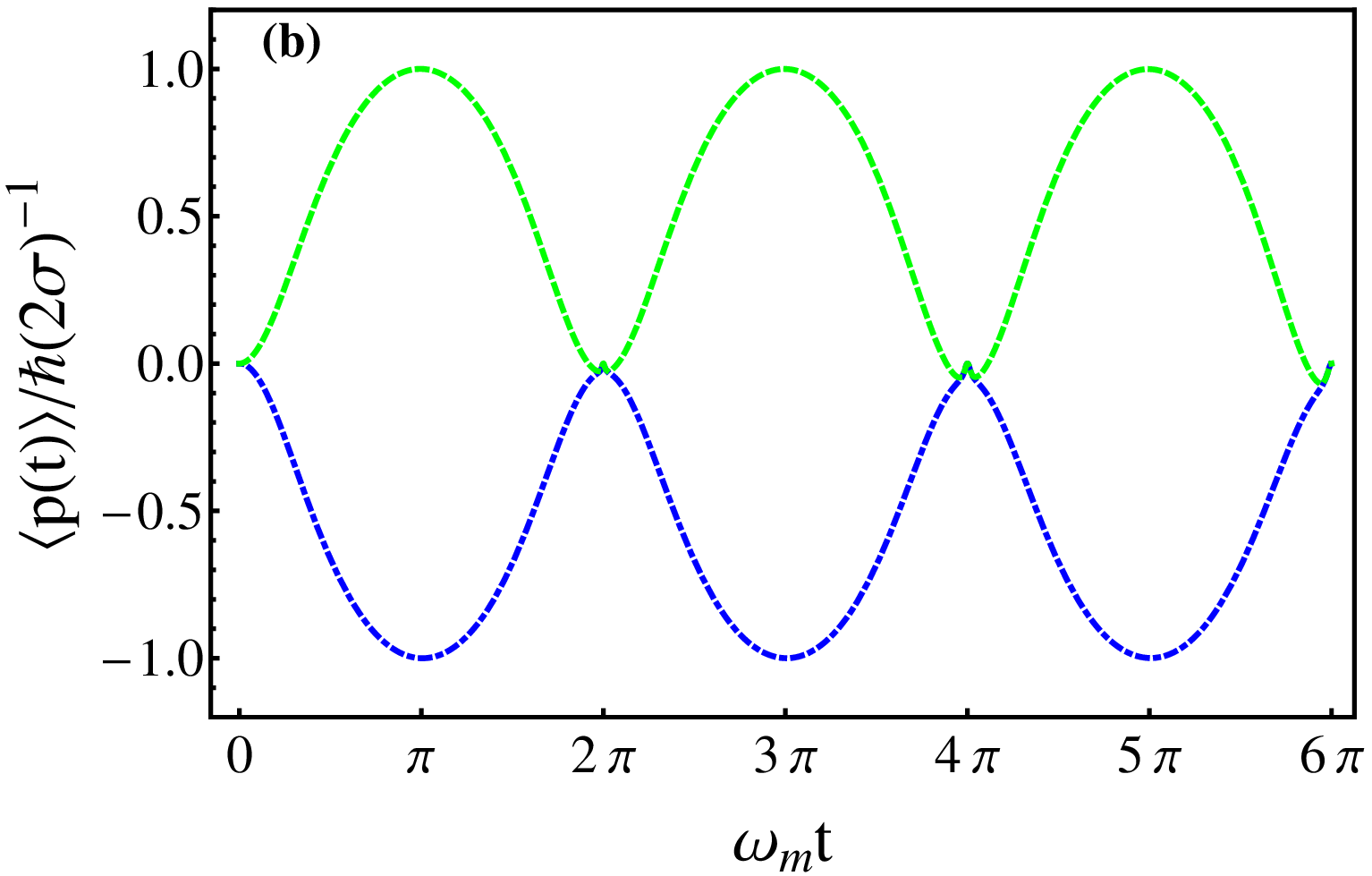}}
\end{tabular}
\caption{(a) The average displacement $\langle p(t)\rangle/\hbar(2\protect%
\sigma)^{-1}$ with $k=0.005$ for $|\protect\alpha|=4$, $\protect\theta=0$
(black-dotted line) and $\protect\theta=\protect\pi $ (red-solid line). (b)
The average displacement $\langle p(t)\rangle/\hbar(2\protect\sigma)^{-1}$
with $k=0.005$ for $|\protect\alpha|=1/2$, $\protect\theta=\protect\pi/2$
(blue-dotted-dashed line) and $\protect\theta=3\protect\pi/2$ (green-dashed
line)}
\end{figure}
Substituting Eq. (\ref{jj}) into Eq. (\ref{bbbb}) using $p$ instead of $q$,
then we have
\begin{align}
\langle p(t)\rangle & =-i\frac{\hbar }{2\sigma }[\varphi _{1}(\gamma
,t)-\varphi _{1}^{\ast }(\gamma ,t)-e^{-\frac{|\varphi _{1}(\gamma
,t)-\varphi _{0}(\gamma ,t)|^{2}}{2}}  \nonumber \\
& (e^{i\phi (t)+\phi (\alpha ,t)+i\tau -D(\gamma ,t)}(\varphi _{1}(\gamma
,t)-\varphi _{0}^{\ast }(\gamma ,t))  \nonumber\\
& +e^{-(i\phi (t)+\phi (\alpha ,t)+i\tau )-D(\gamma ,t)}(\varphi _{1}^{\ast
}(\gamma ,t)-\varphi _{0}(\gamma ,t)))  \nonumber \\
& +\varphi _{0}(\gamma ,t)-\varphi _{0}^{\ast }(\gamma ,t)]/[2-e^{-\frac{%
|\varphi _{1}(\gamma ,t)-\varphi _{0}(\gamma ,t)|^{2}}{2}}  \nonumber \\
& (e^{i(\phi (\alpha ,t)+\tau )-D(\gamma ,t)}+e^{-i(\phi (\alpha ,t)+\tau
)-D(\gamma ,t)})].  \label{ll}
\end{align}

The average displacement $\langle p(t)\rangle/\frac{\hbar}{2\sigma}$ of the
mirror momentum is shown in Fig. 4(b) as a function of time $\omega_{m}t$
with $k=0.005$, $\gamma=0$ (solid line) and $\gamma=0.005$ (dashed line). It
can be seen that the maximal amplifications occur at time $\omega
_{m}t=(2n+1)\pi$ $(n=0,1,2,\cdots)$ and can reach the level of the ground
state fluctuation \cite{Marshall03} $\langle p\rangle$ $=-\frac{\hbar}{%
2\sigma}$. So the amplification of the displacement of the mirror momentum
caused by one photon can be detected in principle. We can also see from Fig.
4(b) that all the amplification values are reduced (dashed line) in the
presence of damping. But because the actual damping $\gamma$ ($%
\gamma=5\times10^{-7}$ in \cite{Bouwmeester12}) in optomechanical cavity is
very small, so the amplifications is barely affected by the damping.

Based on Eq. (\ref{ggg}), we can also make a small quantity expansion about
time $\omega _{m}t=\pi $ till the second order. Suppose that $|\omega
_{m}t-\pi|\ll 1$, and $k\ll 1$, then
\begin{eqnarray}
|\chi _{os}(t)\rangle &\approx &\frac{1}{2}[2k|1\rangle _{m}+i2k|\alpha
|(2\sin \theta -(\omega _{m}t-\pi )  \nonumber \\
&&\cos \theta ))|0\rangle _{m}],  \label{mm}
\end{eqnarray}%
where the first term of Eq. (\ref{mm}) are generated by the amplitude $\xi
(t)$ of coherent state $|\xi (t)\rangle $, while the second term is due to
the relative phase $e^{i\phi (\alpha ,t)}$ caused by the noncommutativity of
quantum mechanics. Based on Eq. (\ref{mm}), when
\begin{equation}
\pm 2k=2k|\alpha |(2\sin \theta -(\omega _{m}t-\pi )\cos \theta ),
\label{nn}
\end{equation}%
we can obtain the maximal value $\frac{\hbar }{2\sigma }$ and minimal value $%
-\frac{\hbar }{2\sigma }$, respectively. So the key to understand the
amplification is the superposition of the vacuum state and one phonon state
of the mirror, which is due to the relative phase $e^{i\phi (\alpha ,t)}$
caused by the noncommutativity of quantum mechanics.
\begin{figure}[tbp]
\centering
{\includegraphics[scale=0.6]{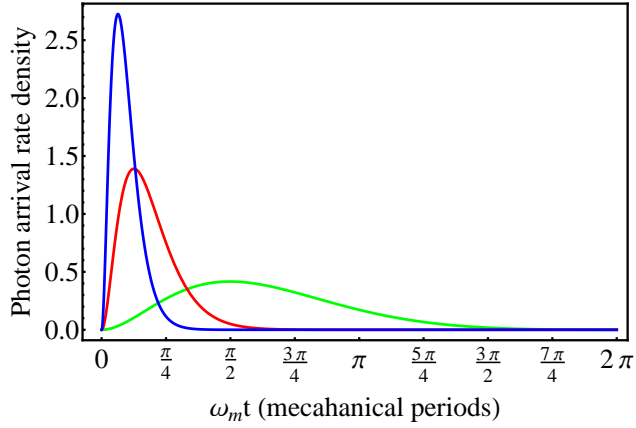}}
\caption{For $|\protect\alpha|=1/2$ and $\protect\theta=0$ with $k=0.005$,
Photon arrival probability density vs arrival time for a successful
postselection with $\protect\kappa=\protect\omega_{m}$ (green line), $%
\protect\kappa=5\protect\omega_{m}$ (red line) and $\protect\kappa=10\protect%
\omega_{m}$ (blue line)}
\end{figure}

Fig. 5(a) shows $\langle p(t)\rangle /\frac{\hbar }{2\sigma }$ for the
amplitude $|\alpha |=4$, the phase $\theta =0$ (black-dotted line) and $\pi $
(red-solid line). We can see from Fig. 5(a) that the maximal amplification
value occur around time $\omega _{m}t=(2n+1)\pi $ $(n=0,1,2,\cdots )$. It is
clear that the maximal amplification around time $\omega _{m}t=(2n+1)\pi $ $%
(n=0,1,2,\cdots )$ can be explained by the expression of Eq. (\ref{nn}).
Fig. 5(b) shows $\langle p(t)\rangle /\frac{\hbar }{2\sigma }$ for the
amplitude $|\alpha |=1/2$, the phase $\theta =\pi /2$ (black-dotted line)
and $3\pi /2$ (red-solid line). In Fig. 5(b), we find that the maximal
amplification value occur at time $\omega _{m}t=(2n+1)\pi $ $(n=0,1,2,\cdots
)$. In the same way, it can also be explained by the result of Eq. (\ref{nn}%
). In one word, the maximal amplification is caused by the equal
superposition of $|0\rangle $ and $|1\rangle $, due to the presence of the
relative phase caused by the noncommutativity of quantum mechanics.

\subsection{ Discussion}

For the feasibility of the proposed scheme, we consider the experimental
experiments from two aspects. First, the mechanical oscillator (mirror) of
our device is initially prepared in coherent state. Coherent state of the
mechanical oscillator has been prepared with itinerant microwave fields \cite%
{Lehnert13}.

We shall show that how the photon arrival rate density varies with time
after the single photon is detected at the dark port. The probability
density of a photon being released from the optomechanical cavity after time
$\omega _{m}t$ is
\begin{equation}
\kappa \exp (-\kappa t),  \label{ppo}
\end{equation}%
where $\kappa $ is the decay rate of the cavity. The probability of a
successful post-selection being released after $\omega _{m}t$ is $\frac{1}{2}%
(1-\exp [-\frac{|\xi (t)|^{2}}{2}]\cos [\phi (t)+\phi (\alpha ,t)])$. For $%
k\ll 1$, this is approximately
\begin{equation}
\frac{1}{4}[|\xi (t)|^{2}+\phi (\alpha ,t)^{2}].  \label{opp}
\end{equation}%
Let us multiply Eq. (\ref{ppo}) and Eq. (\ref{opp}), then we obtain the
photon arrival rate density in optomecanical cavity
\begin{equation}
\frac{\kappa }{4P}\exp (-\kappa t)(|\xi (t)|^{2}+\phi (\alpha ,t)^{2}),
\label{pop}
\end{equation}%
where $P$ is overall single photon probability of the state in Eq. (\ref{ggg}%
):
\begin{eqnarray}
P &=&\frac{1}{4}\int_{0}^{\infty }\kappa \exp (-\kappa t)(|\xi (t)|^{2}+\phi
(\alpha ,t)^{2})dt  \nonumber \\
&=&\frac{k^{2}\omega _{m}^{2}}{2}(2\kappa ^{2}+5\omega _{m}^{2})/(\kappa
^{4}+5\kappa ^{2}\omega _{m}^{2}+4\omega _{m}^{4})  \label{opo}
\end{eqnarray}%
when $|\alpha |=\frac{1}{2}$ and $\theta =0$.

Fig. 6 show that the photon arrival rate density. It can be seen clearly
that in the bad-cavity limit $\kappa >\omega _{m}$ and as the decay rate of
the cavity $\kappa $ increases, such as $\kappa =10\omega _{m}$, the photon
arrival rate density increasingly distributes mainly at time $t$ near $0$
where it is very narrow. Because of the photon arrival rate density
concentrating near the zero time (blue line) in Fig. 6 and the maximal
amplification occurring at time $t$ near $0$ (blue solid line) in Fig. 2,
for a repeated experimental setup with identical conditions, the "average"
position displacement of the pointer is given by

\begin{eqnarray}
\overline{\langle q(t)\rangle } &=&\frac{\kappa }{4P}\int_{0}^{\infty }\exp
(-\kappa t)(|\xi (t)|^{2}+\phi (\alpha ,t)^{2})\langle q(t)\rangle dt  \nonumber\\
&=&0.98\sigma ,  \label{xxy}
\end{eqnarray}%
where $\langle q(t)\rangle $ is the same as $\langle q(t)\rangle $ of Eq. (%
\ref{cccc}). In principle, this result can be experimentally detected since
it is almost close to the strong coupling limit \cite{Marshall03}, i.e., the
level of vacuum-state fluctuation $\sigma $.

The second, for the above analysis, we discuss experimental requirements for
the optomechanical device. Here we need $k$ is high enough so that the
probability of successful postselection is common which depends on the dark
count rate of the detector and the stability of the setup. As shown in Eq. (%
\ref{opo}), the probability of successful postselection in an optomechanical
device with $\kappa =10\omega _{m}$ is approximately $0.01k^{2}$. The window
in which the detectors will need to be open for photons is approximately $%
1/\kappa $, leading to a requirement that the dark count rate be lower than $%
0.01k^{2}\kappa $. Because the best silicon avalanche photodiode have dark
count rate of $\sim 2$ Hz, so we get $k$ $\geq $ $0.0033$ for a $450$ kHz \
device with $\kappa =10\omega _{m}$. Therefore, for Proposed device no. 2
\cite{Bouwmeester12}, we need to change some parameters, including
mechanical frequency $f_{m}=450$ kHz and sideband resolution $\kappa
=10\omega _{m}$, implying that the optical finesse $F$ in Proposed device
no. 2 is reduced to $3.33\times 10^{2}$. Other parameters do not change.
Such an optomechanical cavity is easily to be prepared under the current
conditions. Therefore, the implementation of the scheme provided here is
feasible in experiment.

\section{Conclusion}

In conclusion, we have investigated the weak measurement amplification with
a coherent state pointer. It is regarded as a fire-new mechanism because the
relative phases between the pointer states after the postselection can be
due to the noncommutativity of quantum mechanics, which is different from
the standard weak measurement \cite{Aharonov88,Simon11} where the relative
phases are prepared through the postselection. We find that the maximal
amplification of the displacement of the mirror's position in optomechanical
system can occur near $\omega_{m}t=0$, which can't be achieved if the mirror
is initially prepared in the ground state \cite{Li14}. This result is very
important for bad optomechanical cavities, and because of this, the
implementation of our scheme is feasible in experiment. So these results
extend application of weak measurement in optomechaical system and also
deepen our understanding of the weak measurement.

\section{ACKNOWLEDGMENT}

We thank Tao Wang for helpful discussions. This work was supported by the
National Natural Science Foundation of China under grants No. 11175033. Y.
M. Y. and X. M. L. acknowledge support from the National Natural Science
Foundation of China (Grant Nos. 61275215) and the National Fundamental
Research Program of China (Grant No. 2011CBA00203). \newline
\appendix

\section{ Weak measurement with a ground state pointer}

In Ref. \cite{Simon11}, they consider the standard weak measurement model
considered in Section II but the initial state of the pointer is assumed to
be the ground state $|0\rangle _{m}$. Suppose the state $|+\rangle =\frac{1}{%
\sqrt{2}}(|0\rangle _{s}+|1\rangle _{s})$ is the initial state of the system
to be measured, where $|0\rangle _{s}$ and $|1\rangle _{s}$ is eigenstates
of $\hat{\sigma}_{z}$. According to the Hamiltonian of Eq. (\ref{bb}), then
the time evolution of the total system is given by
\begin{eqnarray}
e^{-\frac{i}{\hbar }\int \hat{H}dt}|+\rangle |0\rangle _{m} &=&\exp [-\eta
\hat{\sigma}_{z}(\hat{c}-\hat{c}^{\dagger })]|+\rangle |0\rangle _{m}  \nonumber\\
&=&\frac{1}{\sqrt{2}}(|0\rangle _{s}D(\eta )|0\rangle _{m}  \nonumber\\
&+&|1\rangle _{s}D(-\eta )|0\rangle _{m}),  \label{ss}
\end{eqnarray}%
where $D(\eta )=\exp [\eta \hat{c}^{\dagger }-\eta ^{\ast }\hat{c}]$ with $%
\eta =\frac{\hbar \chi }{2\sigma }$ is a displacement operator and $\eta \ll
1$. In the weak measurement regime \cite{Aharonov88} the post-selected state
of the system is closely orthogonal to the initial state of the system which
is usually chosen as $\varepsilon |+\rangle +|-\rangle $, where $%
|\varepsilon |\ll 1$. After postselection the final state of the pointer
became
\begin{eqnarray}
|\psi \rangle _{m} &=&\frac{1}{\sqrt{2}}[(1+\varepsilon )D(\eta )|0\rangle
_{m}-(1  \notag \\
&-&\varepsilon )D(-\eta )|0\rangle _{m}].  \label{tt}
\end{eqnarray}%
When $|\varepsilon |\ll 1$ and $\eta \ll 1$, there is
\begin{eqnarray}
|\psi \rangle _{m} &\approx &\frac{1}{2}[(1+\varepsilon )(1-\eta \hat{\sigma}%
_{z}(\hat{c}-\hat{c}^{\dagger }))|0\rangle _{m}  \nonumber \\
&-&(1-\varepsilon )(1-\eta \hat{\sigma}_{z}(\hat{c}-\hat{c}^{\dagger
}))|0\rangle _{m}]  \nonumber \\
&\approx &\varepsilon |0\rangle +\eta |1\rangle .  \label{xx}
\end{eqnarray}%
Noted that the tiny relative phase $\varepsilon $ arise from a
near-orthogonal postselection on the system. Using the expression of the
pointer's displacement
\begin{equation}
\langle \hat{q}\rangle =\frac{\langle \psi |_{m}\hat{q}|\psi \rangle _{m}}{%
\langle \psi |\psi \rangle _{m}}-\langle 0|_{m}\hat{q}|0\rangle _{m},
\label{yy}
\end{equation}%
and
\begin{equation}
\langle \hat{p}\rangle =\frac{\langle \psi |_{m}\hat{p}|\psi \rangle _{m}}{%
\langle \psi |\psi \rangle _{m}}-\langle 0|_{m}\hat{p}|0\rangle _{m}.
\label{zz}
\end{equation}%
Hence in this case of the near-orthogonal postselection, i.e., $\langle
-|+\rangle \neq 0$, we can find that
\begin{equation}
\langle \hat{q}\rangle =\frac{2\varepsilon \eta }{\varepsilon ^{2}+\eta ^{2}}%
\sigma  \label{kw}
\end{equation}%
and
\begin{equation}
\langle \hat{p}\rangle =0.  \label{kv}
\end{equation}%
When $\varepsilon =\pm \eta $ we will have the largest displacement $\pm
\sigma $ in position space and when $\varepsilon =0$, indicating that the
post-selected state of the system is absolutely orthogonal to the initial
state of the system, i.e., $\langle -|+\rangle =0$, the displacement of
pointer position is zero. However, the displacement of the pointer is always
zero in momentum space.


\end{document}